\documentstyle[twoside,fleqn,npb]{article}
\def\plus{{\!\!\: + \!\!\;}}
\def\minus{{\!\!\: - \!\!\;}}
\def\z#1{{\zeta_{#1}}}
\def\ca{{C_A}}
\def\cf{{C_F}}
\def\nf{{n^{}_{\! f}}}
\def\S(#1){{{S}_{#1}}}
\def\Ss(#1,#2){{{S}_{#1,#2}}}
\def\Sss(#1,#2,#3){{{S}_{#1,#2,#3}}}
\def\Ssss(#1,#2,#3,#4){{{S}_{#1,#2,#3,#4}}}
\def\Sssss(#1,#2,#3,#4,#5){{{S}_{#1,#2,#3,#4,#5}}}
\def\Npm{{{\bf N_{\pm}}}}
\def\Npmi{{{\bf N_{\pm i}}}}
\def\Nminus{{{\bf N_{-}}}}
\def\Nplus{{{\bf N_{+}}}}

\newcommand{\beq}{\begin{equation}}
\newcommand{\eeq}{\end{equation}}
\newcommand{\bea}{\begin{eqnarray}}
\newcommand{\eea}{\end{eqnarray}}

\newcommand{\MSb}{$\overline{\mbox{MS}}$}

\def\pqq(#1){p_{\rm{qq}}(#1)}
\def\Li(#1,#2){{\rm{Li}}_{#1}(#2)}
\def\H(#1,#2,#3,#4){{\rm{H}}_{#1,#2,#3}(#4)}

\newcommand{\AmS}{{\protect\the\textfont2
  A\kern-.1667em\lower.5ex\hbox{M}\kern-.125emS}}

\title{First results for three-loop deep-inelastic structure functions in 
QCD}

\author{
J.A.M. Vermaseren\address{NIKHEF Theory Group, \\ P.O. Box 41882, 
1009 DB, Amsterdam}, 
S. Moch\address{Deutsches Elektronensynchrotron DESY \\
Platanenallee 6, D--15738 Zeuthen, Germany}%
\thanks{previous$\,$address:$\,$Institut$\,$f\"ur$\,$Theoretische%
$\,$Teilchenphysik Universit{\"a}t Karlsruhe, D--76128 Karlsruhe, Germany} 
and A. Vogt${}^a$}

\begin{document}

\begin{abstract}
As a first step towards the complete calculation of deep-inelastic scattering
at third order of massless perturbative QCD, we have computed the fermionic 
($\nf$) contributions to the flavour non-singlet structure functions in 
unpolarized electromagnetic scattering. We briefly discuss the approach chosen 
for this calculation, the problems we encountered and the status of the 
project. We show the results for the corresponding anomalous dimension in both
Mellin-$N$ and Bjorken-$x$. Together with the $\nf$ part of $A^{}_3$, our 
calculation facilitates the complete determination of the threshold-resummation 
coefficients $B^{}_2$ and $D_2^{\,\rm DIS}$. The latter quantity vanishes in 
the \MSb~scheme.
\end{abstract}

\maketitle

\section{Introduction}

The computation of the three-loop contributions to the anomalous dimensions is 
needed to complete the next-to-next-to-leading order (NNLO) calculations for 
deep-inelastic scattering (DIS). These in their turn are required for the NNLO 
determination of the parton distribution functions (PDF's) that describe the 
quark and gluon contents of the proton. The accuracy of~the experiments~\cite
{Hagiwara:pw} has become such that this order of perturbation theory is needed 
to match it. In turn, the PDF's at NNLO accuracy are required for precise QCD 
predictions for the future experiments at the LHC.

The method we use for obtaining the three-loop structure functions is based 
on the calculation of all Mellin moments as a function of the Mellin moment 
number $N$. Once either all even or all odd moments are known, we can obtain 
the results in Bjorken-$x$ space by means of an inverse Mellin transformation. 
This method has been employed for structure functions since the early days of 
QCD~\cite{Gross:1973rr,Politzer:1973um}.
Only the coefficient functions are the two-loop level were first obtained by a 
different approach~\cite{vanNeerven:1991nn,Zijlstra:1991qc,Zijlstra:1992kj,%
Zijlstra:1992qd}. However, the $N$-space method was later used to confirm 
those results~\cite{Moch:1999eb}.

Until the calculation of the two-loop anomalous 
dimensions~\cite{Floratos:1977au,Gonzalez-Arroyo:1979df} this method could 
still be applied in a rather direct way. The two-loop computation of 
$\sigma^{}_L/\sigma^{}_T$ required more finesse, and recursion 
relations~\cite{Kazakov:1988jk,Kazakov:1990jm}, also called difference 
equations, entered the scene. The various mixings of the flavour 
contributions were solved in the calculation of a number of fixed moments 
at the three-loop level~\cite{Larin:1994vu,Larin:1997wd,Retey:2000nq}. This 
method turned out to be so powerful that not only the anomalous dimensions 
could be obtained, but also the three-loop coefficient functions as the 
beginning of the NNNLO computation. Hence it seemed best to continue along 
this path and obtain all moments. This avoids the difficult problem of 
operator mixings and gets us the whole three-loop order in one go.

\section{Eight Problems}

In this section we discuss eight of the main problems that were encountered. 
Basically these are problems that occur in any high-order calculation. Their 
solution however depends of the particular calculation.

\vspace*{1mm}
Problem 1

\noindent For the current calculation the mathematics of the answer had to 
be understood better. Hence first the properties of harmonic sums~\cite
{Gonzalez-Arroyo:1979df,Gonzalez-Arroyo:1980he,Vermaseren:1998uu,%
Blumlein:1998if} and harmonic polylogarithms~\cite{Goncharov,Borwein,%
Remiddi:1999ew} had to be studied. In addition procedures were obtained to 
go from one to the other by means of an inverse Mellin transform~\cite
{Remiddi:1999ew,Moch:1999eb}. All this was programmed in FORM~\cite
{Vermaseren:2000nd}. The answer of our calculations will comprise harmonic 
sums of up to weight 6 for the Mellin moments. In Bjorken-$x$ space we will 
obtain harmonic polylogarithms with weights up to 5.

\vspace*{1mm}
Problem 2

\noindent Next we need a scheme in which all integrals are reduced to a 
set of master integrals. To find these reduction algorithms in a way that 
they can handle any integral of the necessary topologies is much work. We 
have to deal with ${\cal O}(70)$ (sub)$\,$topologies. For some the algorithms 
are trivial, but for the most difficult cases it may take a few months to 
derive a good algorithm. In general the equations are based on integration 
by parts~\cite{'tHooft:1972fi,Tkachov:1981wb,Chetyrkin:1981qh,Tkachov:1984xk}, 
scaling equations, form-factor analysis~\cite{Passarino:1979jh} and some 
equations that fall in a special category because they involve a careful study 
of the parton-momentum limit $P\cdot P \rightarrow 0$.

\vspace*{1mm}
Problem 3

\noindent The master integrals can usually be determined by difference 
equations. For the determination of the $\nf$ part of the non-singlet 
structure functions we needed only first-order and second-order difference 
equations. For the complete calculation we have encountered equations up to 
fourth order. These equations are solved with a special FORM program in which 
we substitute a sufficiently general combination of harmonic sums and then 
solve for the (some times thousands of) coefficients. The boundary values are 
provided by the Mincer program~\cite{Larin:1991fz} that was also used for 
obtaining the fixed moments.

\vspace*{1mm}
Problem 4

\noindent The reduction equations have to be derived in a way that avoids 
spurious poles. Some equations may introduce powers of $1/\epsilon\,$ that 
are superfluous. The general rule is that one has to introduce at most one 
power of $1/\epsilon\,$ for each line that is eliminated. If there are more 
poles, the integrals by which they are multiplied will be needed to a 
corresponding number of extra powers of $\epsilon$. This is not always 
feasible. One usually avoids such extra poles by tedious combinations of 
equations. This is the most difficult part of the reduction scheme.

\vspace*{1mm}
Problem 5

\noindent Next the equations have to be programmed in a way that produces 
results. Brute force application leads eventually to programs that run 
unacceptably long or produce intermediate results that exceed the size of 
the available disks. The solution here is a very careful tabulation of 
integrals of lower complexity. For this one has to define a hierarchy of 
complexity. Currently we have already more than 20000 tabulated integrals.

\vspace*{1mm}
Problem 6

\noindent Because the integrals are functions of a parameter and involve 
harmonic sums of weight 6 (of which there are 486 different ones) each 
tabulated integral typically takes about 20 Kbytes. This means that the 
tables are rather large, ${\cal O}(500$ Mbytes). It is not convenient 
to have to compile such an amount of code for each of the about 10000 
diagrams to be computed, even though the FORM compiler is very fast (about 
2 Mbytes per second on a 1.7 GHz Pentium processor). The solution to this 
problem lies in the use of the new tablebase facilities of 
FORM~\cite{Vermaseren:2002xx}.

\vspace*{1mm}
Problem 7

\noindent All topologies and subtopologies have to be programmed, debugged and 
run. Here the old Mincer program~\cite{Larin:1991fz} for the fixed moments 
forms an indispensable tool. At any moment we can replace the Mellin moment $N$ 
by a fixed integer value and continue with the Mincer program. This way we can 
locate errors rather efficiently. Without this method the chances of obtaining 
the correct answer soon would be virtually zero.

\vspace*{1mm}
Problem 8

\noindent Manage FORM~\cite{Vermaseren:2000nd}. Rare bugs occur. Also for a 
problem of this complexity sometimes new features can bring relief. The 
above-mentioned feature of the tablebases is an example.

\section{Status}

On the level of integrals we distinguish basic building blocks (BBB's) and 
composite building blocks (CBB's). The BBB's are integrals in which the 
parton momentum $P$ flows only through a single line in the diagram. These 
have been programmed completely and the vast majority of necessary integrals 
have been tabulated. Occasionally we still have to add some integrals to the 
tables.  
We have reduction schemes for all CBB's. 
They are switched on and debugged one by one at the 
moment. As there exists a rather strict hierarchy in the complexity of 
diagrams the most complicated will be done last, but that does not necessarily 
imply that the most complicated are the most difficult. The easier topologies 
have all been debugged and most of their diagrams have been run. At the moment 
the more complicated topologies are being treated and run.

\section{Some results}

We have completed the diagrams which contribute to the $\nf$ part of the 
non-singlet structure functions. The results for the anomalous dimension and
some parameters of the soft-gluon resummation are presented below. We hope 
to finish the complete non-singlet part early in the 2003. Later that year 
the singlet results should follow. 
All diagrams we have run check with the Mincer results for several fixed 
moments. Our results include both the anomalous dimensions and the coefficient 
functions. For the latter the reader is referred to ref.~\cite{Moch:2002sn}.

The fermionic three-loop contribution to the even-$N$ non-singlet \MSb\ 
anomalous dimension, with the expansion parameter normalized as 
$\alpha_{\rm s}/(4 \pi)$, is given by
\begin{eqnarray}
\label{anom}
  \!\!&\!\!&\!\!\gamma^{\,(2)}_{\,\rm ns}(N) \:\: = \:  
 16\, \* \ca \* \cf \* \nf \* \bigg( 
            {3 \over 2} \* \z3
          - {5 \over 4}
          + {10 \over 9} \* \S(-3)
  \nonumber\\\!\!&\!\!&\!\! \mbox{}
          - {10 \over 9} \* \S(3)
          + {4 \over 3} \* \Ss(1,-2)
          - {2 \over 3} \* \S(-4)
          + 2 \* \Ss(1,1)
          - {25 \over 9} \* \S(2)
  \nonumber\\\!\!&\!\!&\!\! \mbox{}
          + {257 \over 27} \* \S(1)
          - {2 \over 3} \* \Ss(-3,1)
		  - \Nplus \*  \bigg[
            \Ss(2,1)
          \minus {2 \over 3} \* \Ss(3,1)
          \minus {2 \over 3} \* \S(4)
          \bigg]
  \nonumber\\\!\!&\!\!&\!\! \mbox{}
          + (1-\Nplus)  \*  \bigg[
            {23 \over 18} \* \S(3)
          - \S(2)
         \bigg]
       - (\Nminus+\Nplus)  \times
  \nonumber\\\!\!&\!\!&\!\! \mbox{}
          \times\bigg[\Ss(1,1)
          + {1237 \over 216} \* \S(1)
          + {11 \over 18} \* \S(3)
          - {317 \over 108} \* \S(2)
  \nonumber\\\!\!&\!\!&\!\! \mbox{}
          + {16 \over 9} \* \Ss(1,-2)
          - {2 \over 3} \* \Sss(1,-2,1)
          - {1 \over 3} \* \Ss(1,-3)
          - {1 \over 2} \* \Ss(1,3)
  \nonumber\\\!\!&\!\!&\!\! \mbox{}
          - {1 \over 2} \* \Ss(2,1)
          - {1 \over 3} \* \Ss(2,-2)
          + \S(1) \* \z3
          + {1 \over 2} \* \Ss(3,1)
           \bigg]
          \bigg)
  \nonumber\\\!\!&\!\!&\!\! \mbox{}
+ 16\, \* \cf \* \nf^{\!\! 2} \* \bigg( 
            {17 \over 144}
          - {13 \over 27} \* \S(1)
          + {2 \over 9} \* \S(2)
  \nonumber\\\!\!&\!\!&\!\! \mbox{}
		+ (\Nminus + \Nplus) \*  \bigg[
            {2 \over 9} \* \S(1)
          - {11 \over 54} \* \S(2)
          + {1 \over 18} \* \S(3)
          \bigg]
          \bigg)
  \nonumber\\\!\!&\!\!&\!\! \mbox{}
+  16\, \* \cf^{\!\! 2} \* \nf \* \bigg( 
            {23 \over 16}
          - {3 \over 2} \* \z3
          + {4 \over 3} \* \Ss(-3,1)
          - {59 \over 36} \* \S(2)
  \nonumber\\\!\!&\!\!&\!\! \mbox{}
          + {4 \over 3} \* \S(-4)
          - {20 \over 9} \* \S(-3)
          + {20 \over 9} \* \S(1)
          - {8 \over 3} \* \Ss(1,-2)
          - {8 \over 3} \* \Ss(1,1)
  \nonumber\\\!\!&\!\!&\!\! \mbox{}
          - {4 \over 3} \* \Ss(1,2)
       + \Nplus \*  \bigg[
            {25 \over 9} \* \S(3)
          - {4 \over 3} \* \Ss(3,1)
          - {1 \over 3} \* \S(4)
          \bigg]
  \nonumber\\\!\!&\!\!&\!\! \mbox{}
               + (1-\Nplus)  \*  \bigg[
            {67 \over 36} \* \S(2)
          - {4 \over 3} \* \Ss(2,1)
          + {4 \over 3} \* \S(3)
          \bigg]
  \nonumber\\\!\!&\!\!&\!\! \mbox{}
       + (\Nminus+\Nplus)  \*  \bigg[
            \S(1) \* \z3
          - {325 \over 144} \* \S(1)
          - {2 \over 3} \* \Ss(1,-3)
  \nonumber\\\!\!&\!\!&\!\! \mbox{}
          + {32 \over 9} \* \Ss(1,-2)
          - {4 \over 3} \* \Sss(1,-2,1)
          + {4 \over 3} \* \Ss(1,1)
          + {16 \over 9} \* \Ss(1,2)
  \nonumber\\\!\!&\!\!&\!\! \mbox{}
          - {4 \over 3} \* \Ss(1,3)
          + {11 \over 18} \* \S(2)
          - {2 \over 3} \* \Ss(2,-2)
          + {10 \over 9} \* \Ss(2,1)
  \nonumber\\\!\!&\!\!&\!\! \mbox{}
          + {1 \over 2} \* \S(4)
          - {2 \over 3} \* \Ss(2,2)
          - {8 \over 9} \* \S(3)
           \bigg]
          \bigg) \:\: .
\end{eqnarray}
Here we have suppressed the argument $N$ of the harmonic sums and used the
notation
\begin{eqnarray*}
 \ \ \ \ \ \ \ \ \Npm \, f(N) & = & f(N \pm 1) \nonumber \\
 \ \ \ \ \ \ \ \ \Npmi\, f(N) & = & f(N \pm i) \:\: .
\end{eqnarray*}

\noindent The corresponding splitting function, as usual defined with a 
relative sign, reads
\begin{eqnarray}
\label{split}
   \!\!&\!\!&\!\! P^{(2)}_{\rm ns}(x) \: = \:   
  16\, \* \ca \* \cf \* \nf \* \bigg(
         \pqq(x) \*  \bigg[
            {5 \over 9} \* \z2
          - {209 \over 216}
  \nonumber\\\!\!&\!\!&\!\! \mbox{}
          - {3 \over 2} \* \z3
          - {167 \over 108} \* \ln(x)
          + {1 \over 3} \* \ln(x) \* \z2
  \nonumber\\\!\!&\!\!&\!\! \mbox{}
          - {1 \over 4} \* \ln^2(x) \* \ln(1\minus x) \:\:\:
          - {7 \over 12} \* \ln^2(x)
          - {1 \over 18} \* \ln^3(x)
  \nonumber\\\!\!&\!\!&\!\! \mbox{}
          - {1 \over 2} \* \ln(x)\* \Li(2,x)
          + {1 \over 3} \* \Li(3,x)
          \bigg]
  \nonumber\\\!\!&\!\!&\!\! \mbox{}
       + \pqq( - x) \*  \bigg[
            {1 \over 2} \* \z3
          - {5 \over 9} \* \z2
          - {2 \over 3} \* \ln(1\plus x) \* \z2
  \nonumber\\\!\!&\!\!&\!\! \mbox{}
          + {1 \over 6} \* \ln(x) \* \z2
          - {10 \over 9} \* \ln(x) \* \ln(1\plus x)
          + {5 \over 18} \* \ln^2(x)
  \nonumber\\\!\!&\!\!&\!\! \mbox{}
          - {1 \over 6} \* \ln^2(x) \* \ln(1\plus x)
          + {1 \over 18} \* \ln^3(x)
          - {10 \over 9} \* \Li(2, - x)
  \nonumber\\\!\!&\!\!&\!\! \mbox{}
          - {1 \over 3} \* \Li(3, - x)
          - {1 \over 3} \* \Li(3,x)
          + {2 \over 3} \* \H(-1,0,1,x)
          \bigg]
  \nonumber\\\!\!&\!\!&\!\! \mbox{}
       + (1+x) \*  \bigg[
            {1 \over 6} \* \z2
          + {1 \over 2} \* \ln(x)
          - {1 \over 2} \* \Li(2,x)
  \nonumber\\\!\!&\!\!&\!\! \mbox{}
          - {2 \over 3} \* \Li(2, - x)
          - {2 \over 3} \* \ln(x)\* \ln(1\plus x)
          + {1 \over 24} \* \ln^2(x)
          \bigg]
  \nonumber\\\!\!&\!\!&\!\! \mbox{}
       + (1-x) \*  \bigg[
            {1 \over 3} \* \z2
          - {257 \over 54}
          + \ln(1\minus x)
          - {17 \over 9} \* \ln(x)
  \nonumber\\\!\!&\!\!&\!\! \mbox{}
          - {1 \over 24} \* \ln^2(x)
          \bigg]
       + \delta(1\minus x) \* \bigg[
            {5 \over 4}
          - {167 \over 54} \* \z2
          + {1 \over 20} \* \z2^2
  \nonumber\\\!\!&\!\!&\!\! \mbox{}
          + {25 \over 18} \* \z3
          \bigg]
          \bigg)
+ 16\, \* \cf \* \nf^{\!\! 2} \* \bigg(
         \pqq(x) \*  \bigg[
            {5 \over 54} \* \ln(x)
  \nonumber\\\!\!&\!\!&\!\! \mbox{}
          - {1 \over 54}
          + {1 \over 36} \* \ln^2(x)
          \bigg]
       + (1-x) \*  \bigg[
            {13 \over 54}
          + {1 \over 9} \* \ln(x)
          \bigg]
  \nonumber\\\!\!&\!\!&\!\! \mbox{}
       - \delta(1\minus x) \* \bigg[
            {17 \over 144}
          - {5 \over 27} \* \z2
          + {1 \over 9} \* \z3
          \bigg]
          \bigg)
  \nonumber\\\!\!&\!\!&\!\! \mbox{}
+ 16\, \* \cf^{\!\! 2} \* \nf \*\bigg(
         \pqq(x) \*  \bigg[
            {5 \over 3} \* \z3
          - {55 \over 48}
  \nonumber\\\!\!&\!\!&\!\! \mbox{}
          + {5 \over 24} \* \ln(x)
          + {1 \over 3} \* \ln(x) \* \z2
          + {10 \over 9} \* \ln(x) \* \ln(1\minus x)
  \nonumber\\\!\!&\!\!&\!\! \mbox{}
          + {1 \over 4} \* \ln^2(x)
          + {2 \over 3} \* \ln^2(x) \* \ln(1\minus x)
  \nonumber\\\!\!&\!\!&\!\! \mbox{}
          + {2 \over 3} \* \ln(x) \* \Li(2,x)
          - {2 \over 3} \* \Li(3,x)
          - {1 \over 18} \* \ln^3(x)
          \bigg]
  \nonumber\\\!\!&\!\!&\!\! \mbox{}
       + \pqq(\minus x) \*  \bigg[
            {10 \over 9} \* \z2
          - \z3
          + {4 \over 3} \* \ln(1\plus x) \* \z2
  \nonumber\\\!\!&\!\!&\!\! \mbox{}
          - {1 \over 3} \* \ln(x) \* \z2
          - {5 \over 9} \* \ln^2(x)
          + {20 \over 9} \* \ln(x)\* \ln(1\plus x)
  \nonumber\\\!\!&\!\!&\!\! \mbox{}
          - {1 \over 9}\* \ln^3(x)
          + {1 \over 3} \* \ln^2(x)\* \ln(1\plus x)
          + {20 \over 9} \* \Li(2,\minus x)
  \nonumber\\\!\!&\!\!&\!\! \mbox{}
          + {2 \over 3} \* \Li(3,\minus x)
          + {2 \over 3} \* \Li(3,x)
          - {4 \over 3} \* \H(-1,0,1,x)
          \bigg]
  \nonumber\\\!\!&\!\!&\!\! \mbox{}
       + (1+x) \*  \bigg[
            {7 \over 36} \* \ln^2(x)
          - {67 \over 72} \* \ln(x)
  \nonumber\\\!\!&\!\!&\!\! \mbox{}
          + {4 \over 3} \* \ln(x)\* \ln(1\plus x)
          + {1 \over 12} \* \ln^3(x)
          + {2 \over 3} \* \Li(2,x)
  \nonumber\\\!\!&\!\!&\!\! \mbox{}
          + {4 \over 3} \* \Li(2,\minus x)
          \bigg]
       + (1-x) \*  \bigg[
            {1 \over 9} \* \ln(x)
          - {10 \over 9}
  \nonumber\\\!\!&\!\!&\!\! \mbox{}
          - {4 \over 3} \* \ln(1\minus x)
          + {2 \over 3} \* \ln(x)\* \ln(1\minus x)
          - {1 \over 3} \* \ln^2(x)
          \bigg]
  \nonumber\\\!\!&\!\!&\!\! \mbox{}
       - \delta(1\minus x) \* \bigg[
            {23 \over 16}
          - {5 \over 12} \* \z2
          - {29 \over 30} \* \z2^2
          + {17 \over 6} \* \z3
          \bigg]
          \bigg)
\end{eqnarray}
where we have used 
\begin{eqnarray*}
 \ \ \ \ \ \ \ \ p_{\rm{qq}}(x) & = & 2\, (1 - x)^{-1} - 1 - x
\end{eqnarray*}
\noindent
and all divergences for $x \!\to\! 1 $ are understood in the sense of 
$+$-distributions.
The $\nf^{\!\! 2}$ part of eqs.~(\ref{anom}) and eqs.~(\ref{split}) has been 
derived before by Gracey~\cite{Gracey:1994nn} and we agree with his result.

Our results also facilitate the determination of coefficients governing 
the soft-gluon (threshold) resummation~\cite{Sterman:1987aj,Catani:1989ne,%
Catani:1991rp} at next-to-next-to-leading logarithmic accuracy~\cite
{Vogt:2000ci}. 

The coefficient $A_3$ arising from initial-state collinear emissions is the 
coefficient of $1/(1\minus x)_+$ in $P_{\,\rm ns}^{(2)}(x)$. Its fermionic part reads  
\begin{eqnarray}
\label{A3}
   A_3\Big|_{\,\nf}\!\!  &\!\! = \!\! &
   C_A C_F \nf \left[ - \frac{836}{27} \!+\! \frac{160}{9}\:\z2
              \! -\! \frac{112}{3}\:\z3 \right]
  \nonumber \\ &\!\! + \!\! & \,
  C_F^{\,2} \nf\, \left[ -  \frac{110}{3}  + 32\:\z3 \right]
  \nonumber \\ &\!\! + \!\! & \,
   C_F \nf^{\!\!2} \left[ - \frac{16}{27}\,\right] \:\: .
\end{eqnarray}
Simultaneously to our work Carola Berger~\cite{Berger:2002cb} has used a method 
based on eikonal expansions to obtain this coefficient. The calculations have 
been performed independently and put on the internet before comparison. The 
result agrees.

From the coefficient of $\ln (1\minus x)/(1\minus x)_+$ in the $\nf$ part of 
the three-loop coefficient function we can furthermore  determine the 
(complete) resummation coefficients $B_2$ and $D_2^{\,\rm DIS}$ which are due
to final-state collinear and large-angle soft gluons, respectively. Previously 
only the sum of these two 
parameters had been determined~\cite{Vogt:2000ci} from the two-loop coefficient 
function of ref.~\cite{vanNeerven:1991nn}. Our result involves a different 
linear combination (with a prefactor $\beta_0$), and hence we can extract both 
individually, yielding
\begin{eqnarray}
  B_2\ \  &\!\! = \!\! &
  C_F^{\,2}  \left[ -\frac{3}{2} - 24\,\zeta_3 + 12\,\zeta_2 \right]
		\nonumber \\ & \!\! + \!\! &
  C_F C_A  \left[ - \frac{3155}{54} + 40\,\zeta_3
                 + \frac{44}{3}\,\zeta_2 \right] 
  \nonumber \\ &\!\! + \!\! & \:
  C_F \nf\,  \left[ \frac{247}{27} - \frac{8}{3}\,\zeta_2 \right] 
  \\
  D_2^{\,\rm DIS}\!\! &\!\! = \!\! & 0 \:\: .
\end{eqnarray}
The last result is intriguing and calls for further studies.

\end{document}